\documentclass[prb,aps,twocolumn,superscriptaddress,showpacs]{revtex4}

\usepackage{graphicx}

\begin{document}

\title{Interplay between static and dynamic polar correlations in relaxor Pb(Mg$_{1/3}$Nb$_{2/3}$)O$_{3}$}

\author{C. Stock}
\affiliation{ISIS Facility, Rutherford Appleton Labs, Chilton, Didcot, OX11 0QX, UK}

\author{L. Van Eijck}
\affiliation{Institut Laue-Langevin, 6 rue Jules Horowitz, Boite Postale 156, 38042 Grenoble Cedex 9, France}

\author{P. Fouquet}
\affiliation{Institut Laue-Langevin, 6 rue Jules Horowitz, Boite Postale 156, 38042 Grenoble Cedex 9, France}

\author{M. Maccarini}
\affiliation{Institut Laue-Langevin, 6 rue Jules Horowitz, Boite Postale 156, 38042 Grenoble Cedex 9, France}

\author{P.M. Gehring}
\affiliation{NIST Center for Neutron Research, NIST, Gaithersburg, Maryland 20899-6100, USA}

\author{Guangyong Xu}
\affiliation{Condensed Matter Physics and Materials Science Department, Brookhaven National Laboratory, Upton, New York  11973-5000}

\author{H. Luo}
\affiliation{Shanghai Institute of Ceramics, Chinese Academy of Sciences, Shanghai, China  201800}

\author{X. Zhao}
\affiliation{Shanghai Institute of Ceramics, Chinese Academy of Sciences, Shanghai, China  201800}

\author{J.-F. Li}
\affiliation{Department of Materials Science and Engineering, Virginia Tech., Blacksburg, Virginia 24061}

\author{D. Viehland}
\affiliation{Department of Materials Science and Engineering, Virginia Tech., Blacksburg, Virginia 24061}

\date{\today}

\begin{abstract}

We have characterized the dynamics of the polar nanoregions in Pb(Mg$_{1/3}$Nb$_{2/3}$)O$_{3}$ (PMN) through high-resolution neutron backscattering and spin-echo measurements of the diffuse scattering cross section.  We find that the diffuse scattering intensity consists of \emph{both} static and dynamic components.  The static component first appears at the Curie temperature $\Theta \sim 400$\,K, while the dynamic component freezes completely at the temperature T$_{f} \sim 200$\,K; together, these components account for all of the observed spectral weight contributing to the diffuse scattering cross section.  The integrated intensity of the dynamic component peaks near the temperature at which the frequency-dependent dielectric constant reaches a maximum (T$_{max}$) when measured at 1\,GHz, i.\ e.\ on a timescale of $\sim 1$\,ns.  Our neutron scattering results can thus be directly related to dielectric and infra-red measurements of the polar nanoregions.  Finally, the global temperature dependence of the diffuse scattering can be understood in terms of just two temperature scales, which is consistent with random field models.

\end{abstract}


\maketitle

\section{Introduction}

\begin{figure}[t]
\includegraphics[width=9cm] {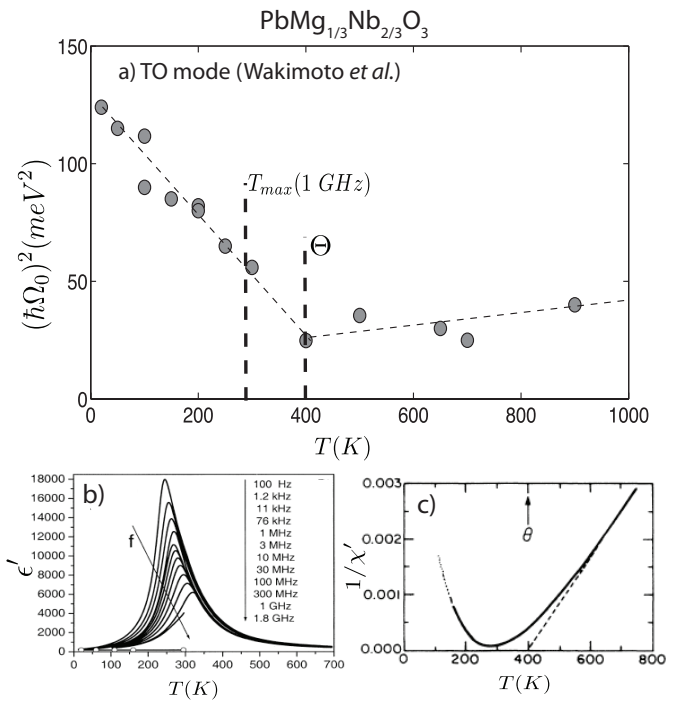}
\caption{$a)$ The square of the soft, transverse optic mode frequency is shown as a function of temperature (data taken from Refs. \onlinecite{Waki02:65} and \onlinecite{Stock05:74}. $b)$ The frequency dependence of the dielectric constant of PMN (data taken from Ref.~\onlinecite{Bovtun04:298}. $c)$ The inverse of the 100-kHz dielectric susceptibility as a function of temperature.  The dashed line represents a high temperature fit and the Curie-Weiss temperature ($\Theta$) is indicated.  The data were taken from Ref.~\onlinecite{Viehland92:46}.}
\label{figure1}
\end{figure}

Lead-based relaxor ferroelectrics are compounds with the general formula PbBO$_{3}$ in which the B-site is disordered.  They exhibit exceptionally large piezoelectric coefficients ($d_{33} \sim 2,500$\,pC/N) and dielectric constants ($\epsilon \sim 25,000$) making them attractive for device applications.~\cite{Ye_rev:98,Park97:82,Xu_rev:xx}  PbMg$_{1/3}$Nb$_{2/3}$O$_{3}$ (PMN) and PbZn$_{1/3}$Nb$_{2/3}$O$_{3}$ (PZN) are prototypical relaxor ferroelectrics and the most studied; both display a broad and unusually frequency-dependent zero-field dielectric response (see Fig.~\ref{figure1}$b$) that contrasts with the sharp and (comparatively) frequency-independent peaks observed in conventional ferroelectrics such as PbTiO$_{3}$ (PT) and BaTiO$_{3}$.

\begin{figure}[t]
\includegraphics[width=8cm] {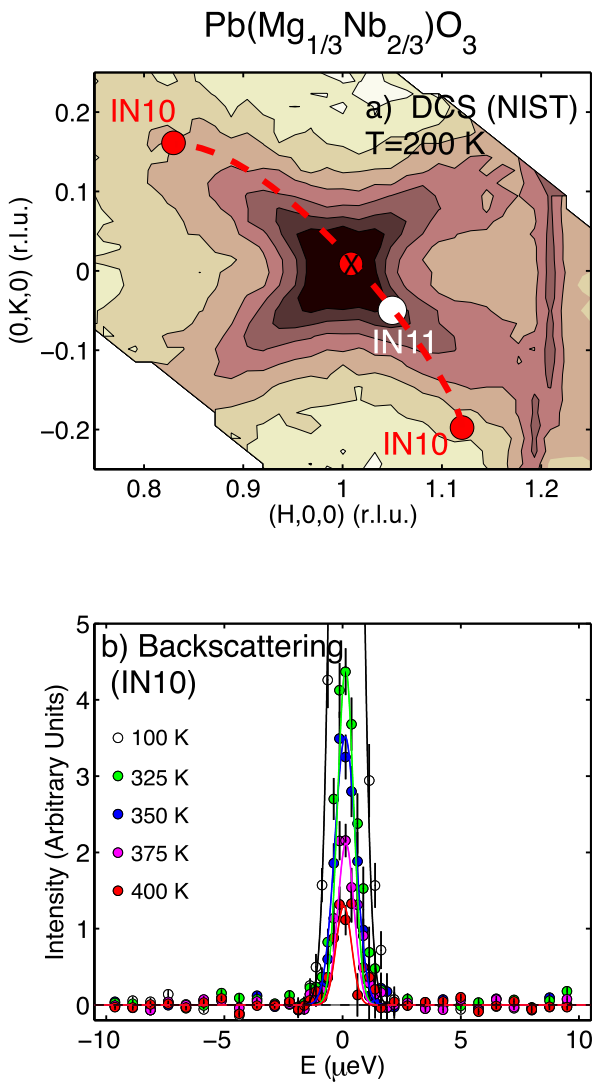}
\caption{ $a)$ Diffuse scattering intensity contours measured near $\vec{Q}$=(100) on the DCS spectrometer located at NIST; the circles illustrate where in reciprocal space the measurements on IN10 and IN11 were conducted.  $b)$ Backscattering spectra measured at several temperatures and summed over the Q-positions indicated in red.  The error bars are related to the square root of the total number of neutron counts.}
\label{diffuse_figure}
\end{figure}

This anomalous dielectric behavior is matched by the odd structural properties of the lead-based relaxors.~\cite{Xu06:79}  Instead of a well-defined structural transition to a long-range ordered ferroelectric ground state, which normally characterizes a typical ferroelectric, lead-based relaxors develop short-range ferroelectric correlations on cooling that are consistent with tiny domains of ferroelectric order embedded within a paraelectric matrix, while the average structural unit cell remains cubic.  These local regions of ferroelectric order, now known as polar nanoregions or PNR, were first postulated to explain the temperature dependence of the optical index of refraction of a variety of disordered ferroelectric materials.~\cite{Burns83:48}  The existence of PNR has since been confirmed by numerous x-ray and neutron scattering studies, which report the presence of strong diffuse scattering at low temperatures.~\cite{You97:79, Hirota02:65,Vak95:37,Vak89:90}  The diffuse scattering in PMN, for example, was recently investigated in great detail and shown to vanish above $\sim 420$\,K.~\cite{Gehring09:79}  Typical diffuse scattering intensity contours measured at 200\,K are displayed in Fig.~\ref{diffuse_figure}$a)$~\cite{Xu_TOF} and show that the neutron diffuse scattering cross section is very broad in momentum, which implies the existence of short-range structural correlations.  By comparison, the sharp, resolution-limited Bragg peaks that accompany a transition to a normal ferroelectric ground state such as in PbTiO$_{3}$ are indicative of long-range structural correlations.  While the lead-based relaxor diffuse scattering cross section has been modeled extensively, considerable debate persists over the underlying physical origin of the intriguing butterfly-shaped contours illustrated in Fig.\,\ref{diffuse_figure}$a$).~\cite{Vak05:7,Pasciak07:76,Welberry05:38,Welberry06:74,Xu04:70}

Despite uncertainties in the origin of the diffuse scattering cross section, the polar nature of the diffuse scattering in the lead-based relaxors has been well established through three different means.  First, the onset of the diffuse scattering in PMN was shown to coincide with the temperature at which a ferroelectric-active, soft, transverse optic mode reaches a minimum frequency (see Fig.\,\ref{figure1}$a$), and which also coincides with the Curie-Weiss temperature derived from high-temperature susceptibility measurements (Fig.~\ref{figure1} $c$).~\cite{Gehring09:79,Waki02:65,Viehland92:46}  Second, the diffuse scattering was shown to respond strongly to an electric field, which in PMN suppresses the diffuse scattering while simultaneously enhancing the Bragg peaks at low temperatures.~\cite{Gehring04:70,Stock07:76,Xu06:74,Xu05:72} Third, on doping PMN with PT both the piezoelectric properties and the total diffuse scattering increase but then drop sharply above PT concentrations near 32\%, at which point the diffuse scattering is replaced by a well-defined structural transition.~\cite{Matsuura06:74,Cao08:78}  While there is a clear connection between the anomalous dielectric properties and the diffuse scattering in lead-based relaxors, the lattice dynamics remain poorly understood, particularly at long timescales.  Specifically, broadband infrared and frequency-dependent dielectric measurements, and the diffuse scattering cross section measured with neutrons and x-rays are difficult to reconcile because the former two techniques generally probe only the momentum response near $Q=0$.  This problem is further complicated by the fact that neutron scattering measurements are typically limited to energy resolutions $\delta E \sim 1$\,THz, and therefore direct comparisons with low-frequency dielectric data have not been possible.

While comparisons of dielectric and infra-red measurements to neutron and x-ray scattering data are difficult to make, data at the high frequency limit of broadband measurements ($\sim 1$\,THz) on thin films have proven to be in excellent agreement with neutron inelastic scattering measurements of the soft transverse optic mode.~\cite{Bovtun04:298}  Low frequency measurements, however, have suggested the presence of significant dynamics that appear to be strongly correlated with the anomalous structural properties.~\cite{Kamba05:17}  While it is tempting to attribute these long timescale dynamics to the diffuse scattering associated with the PNR, several studies using thermal neutron spin-echo and cold neutron triple-axis techniques, which provide excellent energy resolution, have reported the diffuse component to be resolution-limited (and hence static) at all temperatures.~\cite{Vak05:7,Hlinka03:15}  The supposed static nature of the diffuse scattering cross section has led to some models that rely solely on strain or defects to describe the cross section and reconcile it with the dynamics observed in infra-red and dielectric measurements.  On the other hand, other neutron studies have found some evidence of low-energy quasielastic scattering that has been suggested to be correlated with the dielectric properties.~\cite{Hiraka04:70,Gvasaliya05:17,Cowley09:378}  However, because the same low-energy inelastic cross section also contains significant acoustic phonon scattering, the basic nature and existence of any quasielastic component remains unclear.~\cite{Stock05:74}  For this reason it is highly desirable to study the diffuse scattering in lead-based relaxors with an energy resolution that is comparable to that of frequency-dependent dielectric measurements so that a direct comparison between the results from these different techniques can be made.

We have measured the diffuse scattering using cold neutron backscattering and spin-echo techniques in an effort to characterize the dynamics of the PNR in the relaxor PMN.  The large dynamic range offered by both techniques allows a direct comparison with dielectric measurements, which probe dynamics on timescales less than $\sim 1$\,GHz.  While our backscattering data reproduce previously published results concerning the static nature of the diffuse cross section, our spin-echo experiments reveal that the diffuse scattering is in fact described by two components - one static and one dynamic.  The static component is onset at 400\,K, where the ferroelectric-active, soft phonon reaches a minimum frequency.  The temperature dependence of the static component matches well with our backscattering results and with previous data using coarser energy resolution performed on triple-axis spectrometers.~\cite{Hiraka04:70}  The dynamic component is well described by a single relaxational decay time ($\tau$) on the order of $\sim 0.1$\,ns.   The temperature dependence of the relaxational time is described by a simple Arrhenius law with an activation temperature of 1100\,K.  We suggest that T$_{f} \sim 200$\,K, which is the so-called Vogel-Fulcher temperature, is the temperature at which the PNR freeze entirely, and that 400\,K is correlated with the onset of static, short-range ferroelectric correlations (PNR).  The dynamics also reflect the high-temperature scale where previous dielectric and optical index of refraction measurements have suggested the onset of fluctuating polar nanoregions, often referred to as the Burns temperature.

\section{Experimental Details}

Our neutron scattering experiments were conducted on the IN10 backscattering and IN11 spin-echo spectrometers located at the Institut Laue Langevin in Grenoble, France.  IN10 consists of a large, vibrating Si(111) monochromator that Doppler shifts incident neutrons with energies near $E_{i}=2.08$\,meV and directs them on to the sample.  Neutrons scattered from the sample are then energy analyzed by a large bank of Si(111) crystals, which backscatter neutrons having energy $E_f=2.08$\,meV through the sample and onto a series of detectors.  The dynamic range for this spectrometer is $\pm 10$\,$\mu$eV, and the elastic energy resolution is $\delta E=0.5$\,$\mu$eV (half-width).  Neutron spin echo (NSE) spectroscopy differs from other neutron spectroscopic methods in that it measures the real part of the normalized intermediate scattering function $\Re[I(Q,t)/I(Q,0)]$~\cite{Mezei}, where $Q$ is the total momentum transferred to the sample. This is achieved by encoding the neutron's speed into the Larmor precession of its nuclear magnetic moment in a well controlled, externally applied magnetic field. $I(Q,t)$ is the spatial Fourier transform of the Van Hove self correlation function $G(r,t)$ which, essentially, gives the probability of finding a particle after time $t$ at a radius $r$ around its original position~\cite{vanHove}. Furthermore, $I(Q,t)$ is the frequency Fourier transform of the scattering function $S(Q,\omega)$, which is what is measured with neutron backscattering spectroscopy.  We used the NSE spectrometer IN11 in its high resolution set-up ``IN11A''.  The PMN sample is a 3\,cc crystal grown using the Bridgeman technique described elsewhere.~\cite{Luo00:39}  The sample has a room temperature lattice constant of 4.04\,\AA\ and was aligned in the (HK0) scattering plane for all measurements.

\section{Results and Discussion}

We first discuss our measurements of the static diffuse scattering cross section, which is associated with short-range polar correlations.  Fig.~\ref{diffuse_figure}$a)$ shows the geometry of the diffuse scattering intensity contours in PMN measured near $\vec{Q}=(100)$ at 200\,K.  These data were obtained using the Disk Chopper Spectrometer (DCS) located at NIST and have been published and discussed elsewhere.~\cite{Xu_TOF}  The red and white circles indicate the positions in reciprocal space that we studied using the backscattering (IN10) and spin-echo (IN11) techniques, respectively.  Panel $b)$ shows various constant-$Q$ scans measured using IN10.  These data have been corrected for a background measured at 550\,K, above the onset of any diffuse scattering component. The data in panel $b)$ are the sum of the intensities measured at the two $\vec{Q}$-positions indicated by the red circles labeled IN10 in panel $a)$.  As can be seen, both points are located far away from the (100) Bragg peak position and therefore are not contaminated by changes in the Bragg peak intensity as a function of temperature.  The lines in panel $b)$ are fits to a Gaussian with a width fixed to the energy resolution of IN10, which was measured using the incoherent scattering cross section from a Vanadium standard.  The data in panel $b)$ illustrate the onset of the static (resolution-limited) component of the diffuse scattering cross section.  These data also show the absence of any measurable dynamics out to energy transfers of about 10 $\mu$eV between 10\,K and 550\,K (as indicated by the lack of any intensity or temperature dependence).  However the dynamic range probed by backscattering is very limited, and so it is desirable to investigate the inelastic properties of PMN with a technique covering a broader dynamic range while maintaining an excellent energy resolution.  Indeed, it is possible that the spectral weight gathering within the elastic ($E=0$) line at low temperatures may result from a rapid evolution of the dynamics.  To investigate this possibility, we have used spin-echo spectroscopy, which covers a large dynamic range in time simultaneously.

\begin{figure}[t]
\includegraphics[width=8cm] {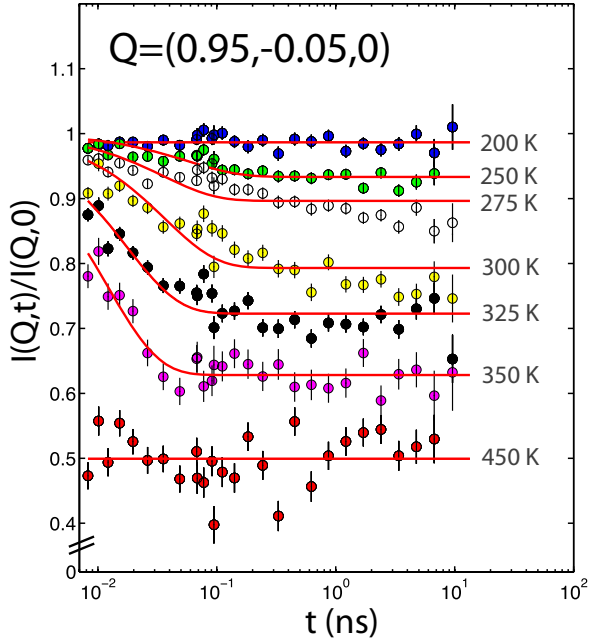}
\caption{Time dependence of the normalized intermediate scattering function $\Re[I(Q,t)/I(Q,0)]$ at various temperatures.  Both dynamic and static components are observed for 200\,K $<T<$ 450\,K, but only the static component is present at 200\,K.  The error bars are related to the square root of the total number of neutron counts.}
\label{NSE_fits}
\end{figure}

The spectra measured using spin-echo (NSE) are plotted in Fig.~\ref{NSE_fits} between 200\,K and 450\,K and characterize the intermediate scattering function $\Re[I(Q,t)/I(Q,0)]$ as a function of $t$.  At 450\,K and above, the data are consistent with a flat line, thus indicating no time dependent dynamics are present within the range accessible with the NSE technique.  The error bars are large for the 450\,K data set because the diffuse scattering intensity is extremely weak at this temperature.  The relatively featureless spectrum at 450\,K contrasts with that measured at 350\,K, which clearly shows a systematic decay with time, thereby illustrating the presence of a dynamic component.  At lower temperatures the decay time increases until at 200\,K only a flat, $t$-independent response is observed, which indicates that the normalized intermediate scattering function is purely static within the limits of the instrumental resolution, i.\ e.\ $\Re[I(Q,t)/I(Q,0)]=1$.  The data in Fig.~\ref{NSE_fits} therefore reflects an interplay between static and dynamic components of the diffuse scattering as the temperature is varied.  Based on the direct connection between the diffuse scattering and the short-range polar correlations observed in PMN, we attribute this interplay to the presence of both static and dynamic polar nanoregions.

To quantify the static and dynamic components we have fit the spectra to a single relaxational form $\Re[I(Q,t)/I(Q,0)]=\alpha+(1-\alpha)e^{-t/\tau}$. The constant $\alpha$ represents the fraction of intensity that is static on the timescale of the measurement, and $\tau$ represents the characteristic decay time of the dynamics.  Based on the fits shown in Fig.~\ref{NSE_fits} we can extract the fraction of the raw uncorrected intensity associated with dynamics ($1-\alpha$) and statics ($\alpha$).  A distribution of decay times can be incorporated into the fit by using the form $\Re[I(Q,t)/I(Q,0)]=\alpha+(1-\alpha)e^{-(t/\tau)^{\beta}}$.  While there is physical justification for a distribution of decay times associated with the dynamics of the polar nanoregions, it is not clear what the value of $\beta$ should be for this case.  Our choice ($\beta=1$) is justified by the fact that our data are well described by a single timescale over a broad temperature range.  We interpret the single decay time extracted from the fits to be the average fluctuation time of the polar nanoregions.

\begin{figure}[t]
\includegraphics[width=8cm] {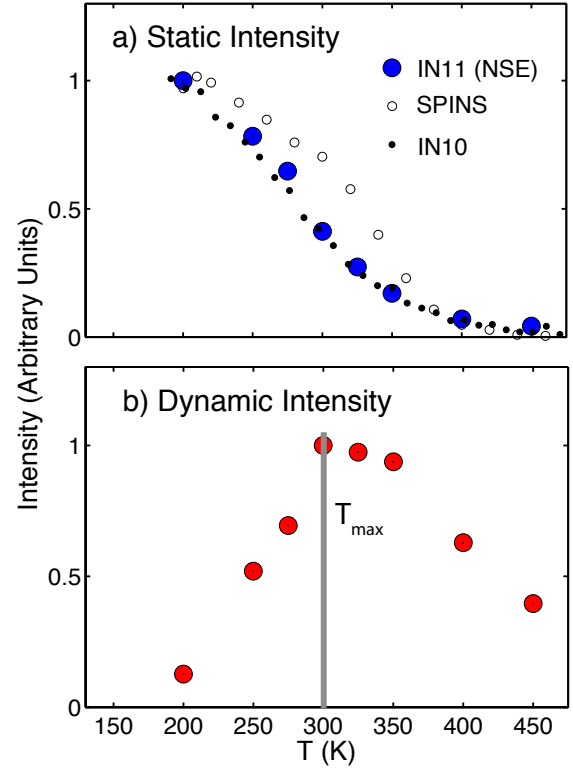}
\caption{ Temperature dependence of the static and dynamic components extracted from fits to the data in Fig.~\ref{NSE_fits}.  $a)$ compares the static intensities measured with SPINS ($\delta E=100$\,$\mu$eV), IN10 ($\delta E=0.5$\,$\mu$eV) and IN11.  Panel $b)$ plots the dynamic component of the intensity as a function of temperature.  The value of $T_{max}$ from Ref.~\onlinecite{Bovtun04:298} measured at 1\,GHz is represented by a vertical line. We emphasize that all data was taken on the same PMN sample.}
\label{intensity}
\end{figure}

The static and dynamic components extracted from Fig.~\ref{NSE_fits} have been corrected for the IN11 instrumental resolution ($I(Q,0)$) and plotted as a function of temperature in Fig.~\ref{intensity}.  Panel $a)$ compares the elastic (or static) intensity measured using three different instruments and experimental resolutions, and panel $b)$ displays the dynamic component normalized to have a maximum value of unity.  The data taken on SPINS, a cold neutron triple-axis spectrometer located at NIST, were measured at $\vec{Q}$=(0.025,0.025,1.05) with an energy resolution of $\delta E=100$\,$\mu$eV.  The backscattering (IN10) data were measured at the $\vec{Q}$ indicated in Fig.~\ref{diffuse_figure}$a)$ and represent the intensity of a Gaussian function of energy fit to the elastic peak.  The NSE data represent the static parameter ($\alpha$) extracted from fits to the NSE spectra described above multiplied by $I(Q,0)$.  Hence the data plotted in Fig. \ref{NSE_fits}$a)$ equals $I(Q,0) \times \alpha$.  All of the data have been normalized to unity at 200\,K.  The static component derived from the NSE fits described above agree well with the static component derived from backscattering.  The data do not agree as well with the intensities derived from poorer energy resolution measurements on SPINS, which used a fixed final energy $E_{f}=4.5$\,meV, and are suggestive of a dynamic component over this region that has been integrated over (i.\ e.\ lumped into the elastic channel) by virtue of the fact that the SPINS instrument provides a substantially poorer energy resolution than do the IN10 backscattering and IN11 NSE spectrometers.

From Fig.~\ref{intensity}$a)$ the onset of the static portion of the diffuse scattering appears between 400\,K and 450\,K.  This is significant when compared to dielectric and other neutron inelastic scattering results.~\cite{Gehring09:79}  The Curie-Weiss temperature $\Theta$ derived from high temperature dielectric data (Fig.~\ref{figure1}$c)$) is 400\,K~\cite{Viehland92:46} and coincides exactly with the temperature at which the ferroelectric-active, soft mode reaches its minimum energy (Fig.~\ref{figure1}$a)$).~\cite{Waki02:65}  The 400\,K temperature scale also matches the temperature at which strong deviations in the coefficient of thermal expansion in PMN are observed.~\cite{Dkhil01:65}  Based on this result we interpret the onset of the static component of the diffuse scattering as the freezing of PNR.  The short-range nature of these frozen regions is evident from the broad nature of the diffuse scattering cross section in momentum space, which was illustrated in Fig.~\ref{diffuse_figure}.

The dynamic component illustrated in panel $b)$ equals $I(Q,0) \times (1-\alpha)$.   On cooling the dynamic component increases, peaks, then decreases; this is consistent with dynamics entering and leaving the time window probed NSE.  The maximum intensity of the dynamic component occurs near $\sim 325$\,K, which is agrees well with the temperature $T_{max}$ at which the peak dielectric response occurs when measured at a frequency of 1\,GHz  (see the vertical line in Fig.~\ref{figure1}$b)$).  We emphasize that $T_{max}$ is not a meaningful temperature scale because it depends on the energy resolution and technique used to measure it.  Rather, $T_{max}$ is a temperature that is characteristic of the dynamics and the frequency (or timescale) at which they are probed.  We also note that contamination of our data from acoustic modes, which have plagued other studies, is unlikely as these phonons reside at much larger energy transfers (or at shorter times $\sim 1$\,THz) than are probed here.  Also, the acoustic phonon intensity increases with temperature as required by the Bose factor; this does not reflect the trend observed in our experiment.

There is little dynamic spectral weight at 200\,K where the normalized NSE spectra (Fig.~\ref{NSE_fits}) nearly reaches unity and becomes flat.  These data can then be understood in terms of two temperature scales.  The first is an upper transition temperature near 400\,K where static, short-range, ferroelectric correlations are onset.  This temperature coincides with the Curie-Weiss temperature derived from high-temperature dielectric data as well as the temperature where the ferroelectric-active, soft transverse optic mode reaches its minimum frequency.  A second important temperature scale exists around 200\,K.  This temperature is defined as that where the short-range polar correlations are truly static.  This temperature is most clearly reflected in electric field studies, which show that the diffuse scattering can be suppressed by an electric field only below $\sim 200$\,K.  This is also the transition temperature $T_{f}$ where dielectric measurements under a poling electric field observe a sharp peak characteristic of the presence of long-range, ferroelectric correlations.

\begin{figure}[t]
\includegraphics[width=8cm] {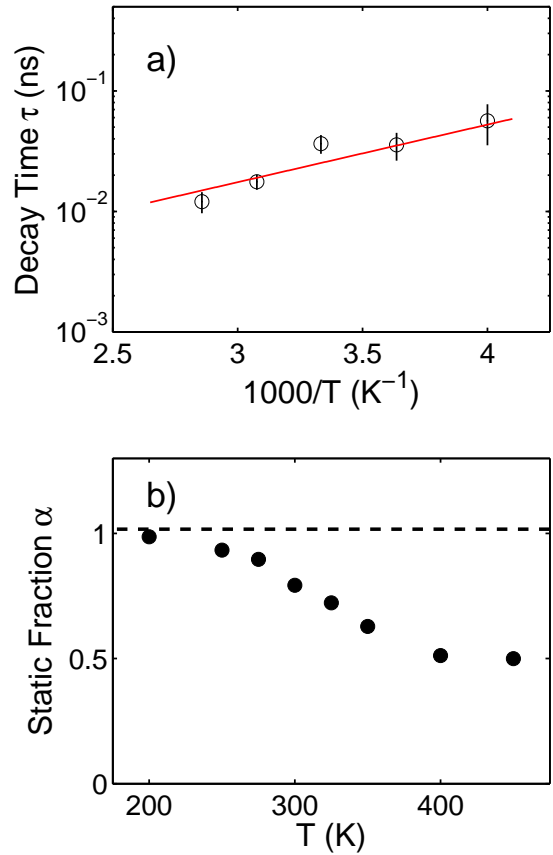}
\caption{$a)$ The decay time $\tau$ is plotted as a function of temperature. The solid line is a fit to an Arrhenius law ($\tau = \tau_{\circ} e^{U/k_{b}T}$) with $U$=1100 K. $b)$ The fitting parameter $\alpha$ is plotted as a function of temperature.}
\label{param}
\end{figure}

We now discuss the temperature dependence of the decay time $\tau$ and the parameter $\alpha$.  Both $\tau$ and the static component of the normalized intermediate scattering function $\Re[I(Q,t)/I(Q,0)]$ are displayed in Fig.~\ref{param}.  Panel $a)$ shows the temperature dependence of $\tau$, which we have chosen to fit to the simple Arrhenius equation $\tau = \tau_{\circ} e^{U/k_{b}T}$ with $U=1100$\,K$\pm 300$\,K.  This value for $U$ was also confirmed through a universal fit of all the data to a single exponential decay following the Arrhenius equation.  Studies of spin-glasses, which exhibit analogous slow dynamics, have often described the observed relaxation rates in terms of a power law or an exponential, Arrhenius-type equation.~\cite{Aeppli85:54,Som82:25}  Regardless of the exact form, all of these equations describe the dynamics in terms of some form of energy barrier and therefore have very similar physical interpretations.  Some of these descriptions for the temperature dependence of the relaxation rate contain a critical temperature as, for example, in fits using a Vogel-Fulcher type analysis.  More complicated fits introducing more fitting parameters beyond the simple Arrhenius law are not justified by our data.  The value of the activation energy $U=1100$\,K$\pm 300$\,K is in reasonable agreement with that ($E_{a} \sim 800$\,K) derived from infrared and infra-red techniques.~\cite{Kamba05:17}  It is also in agreement with the temperature at which the lowest-energy, transverse, optic mode is observed to become soft and broad in energy.~\cite{Gehring01:87}

Panel $b)$ of Fig. \ref{param} displays the value of $\alpha$ obtained from the fits shown in Fig.~\ref{NSE_fits}.  The value of $\alpha$ varies smoothly from 0.5 to 1 with decreasing temperature.  We do not attribute any significance to the high temperature saturation value of 0.5 for two reasons.  First, the normalization value $I(Q,0)$ is based on an energy integration over the time range of the spectrometer.  The value of 0.5 could thus be the result of an incomplete energy integration (and hence determination of $S(Q,0)$) at high temperatures.  This is corroborated by cold neutron spectroscopy data, which provide evidence of the presence of dynamics at high temperatures on the timescale beyond the resolution of our spectrometer.~\cite{Hiraka04:70}  Secondly, when the measured intensity is taken into account, as done in Fig.~\ref{intensity}, the results are consistent with no or little spectral weight residing in the static channel at high temperatures.  The factor of 0.5 does illustrate the presence of very long timescale dynamics at high temperatures.  The dynamics are beyond the time window of the current experiment, but we speculate that the dynamics could result from short-range chemical order found in Ref. \onlinecite{Hiraka04:70}.  Further work involving longer timescales are required to confirm the origin of this extra component.  The temperature dependence of $\alpha$ illustrates that all of the intensity coming from the diffuse scattering below $\sim$ 200 K is static.  This temperature scale for freezing of the polar nanoregions agrees with other techniques.  NMR T$_{2}$ measurements, which probes fluctuations on the order of MHz, have found evidence of an anomaly near this temperature and have attributed it to freezing of the polar correlations.~\cite{Blinc03:91}

Our results contradict claims based on thermal spin-echo measurements presented in Ref.~\onlinecite{Vak05:7} where the diffuse scattering is concluded to be solely static.  Those measurements were conducted on a thermal spin-echo machine with considerably coarser energy resolution and a smaller dynamic range than was used here.  We therefore believe that these previous measurements only probed the static component found in our study.  We note that the results presented here are consistent with data obtained from backscattering, in particular the IN10 data discussed earlier.  Backscattering probes a very narrow dynamic range of $\sim 1$\,ns; by comparison, neutron spin echo covers several orders of magnitude in time.  It is apparent from Fig.~\ref{NSE_fits} that a much broader dynamic range is required to observe both the dynamic and static response.  We conclude then that backscattering is not sensitive to the dynamic component because the dynamic range accessed by that technique is too narrow.

The presence of a dynamic component of the diffuse scattering and a favorable comparison with frequency-dependent dielectric data strongly link the diffuse scattering cross section to the presence of dynamic polar nanoregions.  Our data also clearly illustrate two key temperature scales.  A high temperature scale, where the static component appears ($\sim 400$\,K), and a lower temperature scale ($\sim 200$\,K), where no dynamic component is observed and only static correlations exist.  We do not attribute any physical significance to T$_{max}$, which we believe is associated with the dynamics and not to any transition temperature.   Our interpretation of the upper temperature of 400\,K is also in agreement with the ideas presented in Ref.~\onlinecite{Gehring09:79}, which further offers an alternative understanding of the Burns temperature at 620\,K in terms of dynamics.

Recent reports of a third temperature scale known as $T^{*}$ can also be understood in terms of the dynamics.  Ref.~\onlinecite{Dkhil09:xx} describes a new upper transition based on deviations in the thermal expansion coefficient.  Such a temperature scale may be related to the dynamics and could also be associated with the near-surface effect measured in single crystal samples of the lead-based relaxors PMN and PZN.~\cite{Conlon04:70,Xu06:79}  Indeed, previous studies of the coefficient of thermal expansion in PMN using neutrons have found a strong variation as a function of depth.  While these speculations are inconclusive, further study is required to understand the origin of these upper transition scales.   Dielectric and infra-red work in Ref.~\onlinecite{Toulouse08:369} describe a third temperature scale T$^{*}$, which is 400\,K, and we interpret this temperature as the onset of static polar correlations.  Ref.~\onlinecite{Toulouse08:369} also descrobe a high temperature scale near 600\,K.  This could be related to the dynamics as evidenced by the large activation energy derived in our current experiment.

The interpretation of the dielectric and neutron data in terms of two temperature scales is in agreement with random field models previously proposed to explain the relaxor ferroelectric phase diagram.  References \onlinecite{Stock04:69,Westphal92:68,Fisch03:67} interpret relaxors in terms of a 3D Heisenberg model with cubic anisotropy in a random field imposed by the disorder on the perovskite $B$ site.  In this picture, the true transition temperature would correspond to 400\,K where the soft, transverse optic phonon mode energy reaches a minimum.  However, this model also predicts a second (lower) temperature scale when the energy scale imposed by the cubic anisotropy becomes important.  It was suggested that only below this temperature could long-range, ferroelectric correlations develop in the presence of a random field.  This model is in broad agreement with the data presented here, which can be described using just two temperature scales.

We have presented a high resolution investigation of the diffuse scattering in PMN as a function of temperature that sheds new light on the underlying behavior of the polar nanoregions.  We have proven the existence of a dynamic component to the diffuse scattering and that the temperature dependence is described by an interplay between dynamic and static components.  We find that the static component appears between 400\,K and 450\,K, which is in excellent agreement with the Curie-Weiss temperature derived from dielectric data and also the minimum in the soft mode energy.  From our spin-echo spectra, we have also derived a dynamic fraction and shown that it peaks near the temperature T$_{max}$ found in frequency-dependent dielectric constant measurements.

We would like to thank V. Garcia Sakai, S. Shapiro, J. Gardner and R. Cowley for very helpful discussions.  We would also like to thank P. Phillips at the ISIS Facility for expert technical support.


\end{document}